\documentclass[10pt]{llncs}

\usepackage{endnotes}
\usepackage{times}
\usepackage[T1]{fontenc}
\usepackage{float}
\usepackage{graphicx}
\usepackage{array}
\usepackage{url}
\usepackage{listings}
\usepackage{amsmath,amssymb,latexsym, pifont}
\usepackage[usenames]{color}
\usepackage{ifsym}
\usepackage{wasysym}
\usepackage{booktabs}
\usepackage{mdwlist}
\usepackage{subfigure}
\usepackage[small,labelfont=bf]{caption}
\usepackage{wrapfig}

\let\underscore\_
\newcommand{\myunderscore}{\renewcommand{\_}{\underscore\hspace{0pt}}}
\myunderscore

\usepackage[margin=.70in]{geometry}

\begin{document}
\title{Control Exchange Points: Providing QoS-enabled End-to-End Services\\via SDN-based Inter-domain Routing Orchestration}

\author{Vasileios Kotronis$^1$,
Xenofontas Dimitropoulos$^{2,1}$,
Rowan Kl\"oti$^1$,
Bernhard Ager$^1$,
Panagiotis Georgopoulos$^1$,
Stefan Schmid$^3$\\
\small{$^1$ ETH Zurich, Switzerland \quad\quad $^2$
Foundation of Research and Technology Hellas (FORTH)} \quad\quad
\small{$^3$ T-Labs \& TU Berlin, Germany}
}

\institute{}

\maketitle
















\textbf{Introduction.}
This paper presents the vision of the Control Exchange Point
(CXP) architectural model. The model is motivated by the
inflexibility and ossification of today's inter-domain routing system,
which renders critical QoS-constrained end-to-end (e2e) network services difficult or simply impossible to provide.
CXPs operate on slices of ISP networks and are built on basic Software Defined
Networking (SDN) principles, such as the clean decoupling of the routing
control plane from the data plane and the consequent logical centralization of
control. The main goal of the architectural model is to provide
e2e services with QoS constraints across domains. This is achieved through
defining a new type of business relationship between ISPs, which advertise
partial paths (so-called \emph{pathlets}~\cite{PathletRouting}) with specific properties,
and the orchestrating role of the CXPs, which dynamically stitch them
together and provision e2e QoS. Revenue from value-added services flows from
the clients of the CXP to the ISPs participating in the service.
The novelty of the approach is the combination of SDN programmability
and dynamic path stitching techniques for inter-domain routing, which
extends the value proposition of SDN over multiple domains.
We first describe the challenges related to e2e service provision with the current
inter-domain routing and peering model, and then continue with the benefits of our
approach. Subsequently, we describe the CXP model in detail and report on an initial feasibility analysis.

\textbf{Motivation and Challenges.}
\textit{Complexity and ossification:}
The notorious complexity of the inter-domain routing system
renders its management difficult and error-prone, leading to various
inefficiencies such as suboptimal inter-domain paths. Indicatively,
60\% of all Internet paths today are suffering from triangle
inequality violations~\cite{HotInterconnects}. The current ossification of the
system, hindering the introduction of new solutions,
aggravates the problem further.
Highly popular inter-domain services, such as high-definition e2e
real-time video streaming, already test the limits of the status quo,
or are simply impossible. This is because such services require tight
coordination along entire chains of ISPs demanding QoS
provisioning. More advanced and mission-critical services, such as
telemedical applications, are usually out of the question.


\begin{wrapfigure}{r}{0.4\textwidth}
\vspace{-35pt}
\centering
\includegraphics[width=0.4\textwidth]{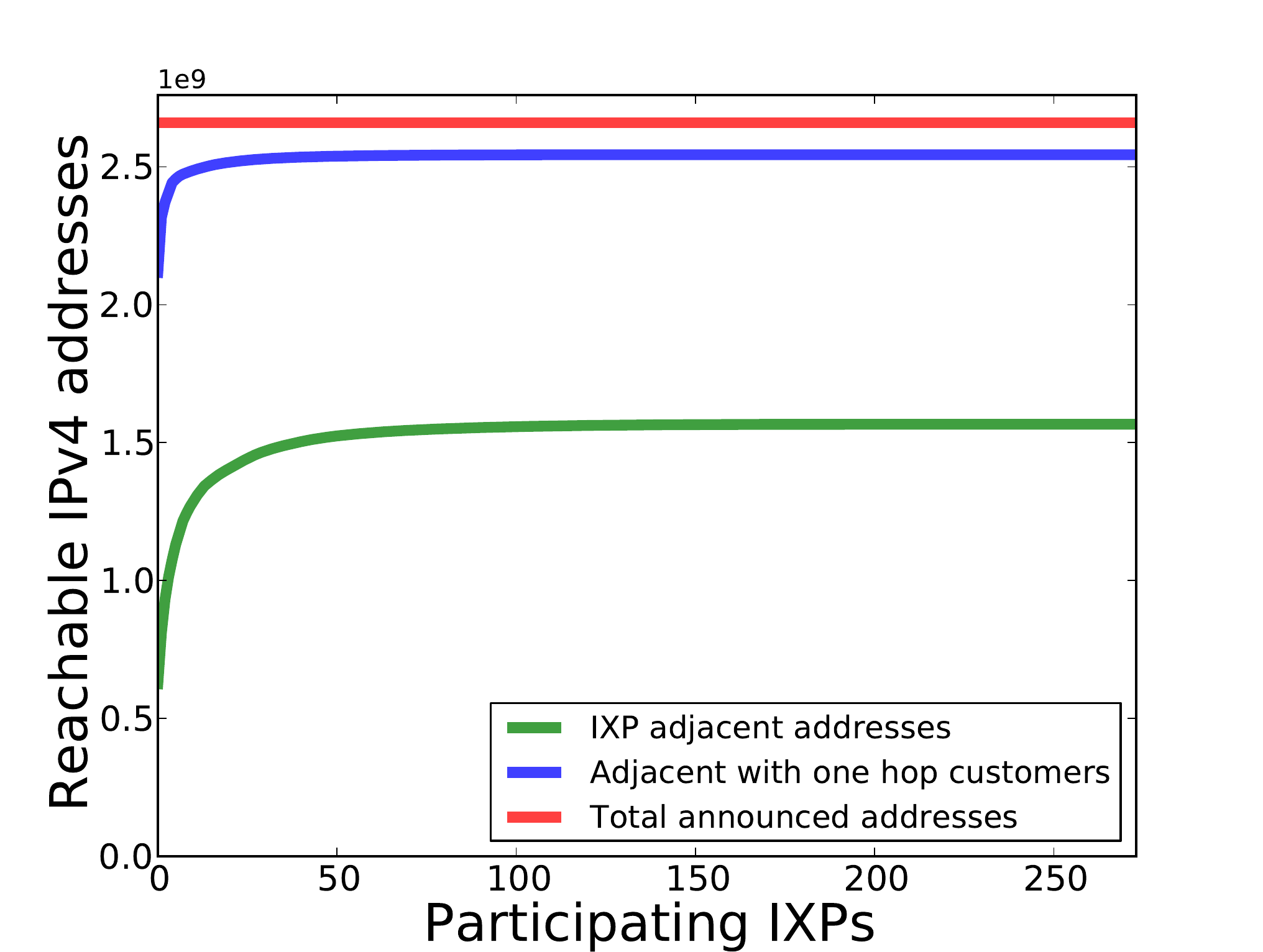}
\vspace{-10pt}
\caption{Cumulative coverage of IP address space originated by IXP members and their 1-hop customer cone versus number of IXPs}
\label{fig:IXP_IP_MUF}
\vspace{-20pt}
\end{wrapfigure}

\textit{Challenges associated with ISP peering:} ISPs today peer with
each other either directly over private peerings, or via the rich IXP ecosystem
that interconnects thousands of ISPs and  is currently morphing the
Internet landscape into a dense mesh,
greatly increasing path diversity. Here we highlight
the limitations of the current peering ecosystem. Firstly, we argue that the
economic model of bilateral business agreements is not suitable for
services offered by multiple ISPs together.
Secondly, peer ISPs exchange all their clients' IP prefixes coarsely, without
any differentiation specific to the cross-domain service that they want
to provide to their clients. Fine-grained service-specific peering is a potential
``nice-to-have'' for ISPs, but BGP does not provide the needed mechanisms
for implementing it with the appropriate granularity (e.g., flow-level).
Thirdly, peering practices today disfavor small (and potentially innovative)
ISPs, since tier-1 ISPs form restricted peering groups and have little incentive to offer
peering, when they could simply charge for transit. This could change if,
for example, a new type of business relationship for cross-domain services
would jointly benefit all chained ISPs, despite their differences in size or their
original business association (customer-provider, peer-peer).


\textit{Benefits of the CXP model:} With this work, we are investigating how SDN
principles can help deal with the inflexibility and suboptimality of
the inter-domain routing system and the issues related to classic ISP
peering. The CXP model enables dynamic
service-specific relationships between ISPs, to provide cross-domain
e2e services that can be guaranteed over the Internet. Example
services that can be provided via peering under the supervision
of the CXP entities include high quality video conferencing,
\emph{telemusic}~\cite{telemusic-LOLA} (teaching music or
      performing with remote participants over long distances by guaranteeing low-delay
      HD video streaming), or mission-critical real time information
      streaming for telemedicine purposes~\cite{telesurgery}: the bandwidth and latency
      sensitive content in this case, could be the live video between an operating
      room in a hospital in Moscow and a doctor performing a remote robot-assisted surgery,
      located in Zurich. Such applications demand a user-transparent, QoS-enabled,
      multi-domain WAN.
\textbf{Control Exchange Points: A new notion of ISP peering based on Software Defined Networking.}
%
A CXP is an external to the ISP entity that orchestrates the e2e
stitching of slices that the ISPs provide, for the benefits of
e2e service revenue. A CXP manages slices of multiple ISPs and
provides inter-domain routing coordination based on SDN APIs. A
slice is defined by a flowspace (associated with a specific
service) and a virtual topology (e.g., pathlets). An ISP abstracts
its network as a set of pathlets connecting the network edges and then advertises
these to the CXP. More specifically, this abstraction could be realized with
tunnels instantiated with e.g., OpenFlow or MPLS. Slices are connected
via inter-domain links e.g., over IXPs, to other ISP domains to form
an inter-domain virtual topology. The pathlet abstraction is bundled with
properties that the ISP provides. For example, a pathlet can be annotated
with minimum bandwidth and/or maximum delay guarantees, the number of routers
it is composed of, the presence of middleboxes associated with a network service, or
disjointness properties related to other offered pathlets. The ISP may choose
to advertise multiple pathlets that connect the same domain edges
(e.g., for backup functionality or even to diversify the provided
guarantees as a form of market segmentation), thus enriching the set of offered pathlets.

The task of the CXP is to admit requests for
QoS-guaranteed e2e paths, embed paths in the inter-domain virtual
topology~\cite{icdcn12}, and monitor the provided QoS guarantees.
The CXP can operate in parallel with BGP, as long as the control over the service-specific
IP flows on the ISP domain edges is outsourced to the CXP, so that the
rest of the BGP prefixes are isolated from the service prefixes. The intra-domain
control of the flows lies, of course, at the ISP's hands. CXPs are the main entities
forming a framework that is based on partial routing outsourcing, in contrast to
full control plane outsourcing~\cite{Hotnets-RaaS}.
In both approaches, inter-domain routing is treated ``as a service'', but the
CXP approach requires much less outsourced control from the IXP's side,
and is more appealing for deployment and market adoption.
Related work~\cite{rexford-RaaS} in routing as a service focuses on similar
notions but neither explores incremental deployment strategies nor
investigates scalable solutions based on SDN mechanisms.

CXPs are elements that operate solely within the control plane.
They interface with ISPs in order to receive pathlet
advertisements and be able to switch (i.e., route) between the ISPs'
pathlets, so as to establish e2e paths dynamically.
Dynamic e2e path control allows to adjust to changing network
conditions and to migrate embedded paths to admit new ones.
We propose that inter-domain connection between ISP slices
takes place over IXPs. The data plane anchors that are used for switching
are network elements deployed within the layer-2 infrastructure of the IXPs.
The main reason for selecting IXPs as the CXP data plane anchors is that
ISPs typically peer with many different IXPs in parallel for traffic
offloading and transit cost reduction. This dense mesh of links provides path diversity
which our model can exploit.
All pathlets are attached to IXPs. The CXP anchors in the IXPs are controlled by the
logically centralized OS platforms~\cite{Hotnets-RaaS} of the CXPs,
based on SDN APIs. The introduction of the SDX~\cite{SDX} concept
and the potential deployment of SDN-enabled infrastructure within IXPs,
complement our model nicely. The CXP could for example interface with the IXP through
the SDX controller APIs and use SDX-defined slices of the current IXP fabric, according
to the service and IXP clientele with which the slices are associated.

\begin{wraptable}{r}{0.35\textwidth}
\vspace{-25pt}
\scalebox{0.68} {
\small
\begin{tabular}{|c|c|c|c|c|c|}\hline 
  & LINX & DE-CIX & Terremark & AMS-IX & Equinix Ashburn\\\hline
LINX & - & 2429 & 1093	& 2443 & 1427\\\hline
DE-CIX & 2429 & - & 1093 & 2429 & 1427\\\hline
Terremark & 1093 & 1093 & - & 1093 & 1093\\\hline
AMS-IX & 2443 &	2429 & 1093	& - & 1427\\\hline
Equinix Ashburn & 1427 & 1427 & 1093 & 1427 & -\\\hline 
\end{tabular}
}
\vspace{-10pt}
\caption{Pairwise path diversity between the 5 largest IXPs}
\label{tab:path-div}
\vspace{-20pt}
\end{wraptable}

Regarding pathlet provision, we propose two models depending on whether
ISPs are willing to provide hard guarantees or not. In the first model, an ISP provides
tunnels across its domain without any strict guarantees. The CXP cleverly
stitches these pathlets together in order to satisfy the e2e requirements of the
end-clients. To do that, active rerouting across unreliable ISP
chains is required, based on real-time measurements taken from the
substrate network. IXPs could ideally serve as monitoring
anchors for such measurements. In the second model, an ISP provides tunnels
with guaranteed performance parameters across its domain.
The CXP is still responsible for stitching these advertised,
locally guaranteed pathlets so as to provide e2e guarantees. In this case,
monitoring is required for verifying that advertised guarantees
are adhered to in practice.
The adoption of the most suitable model for current markets relates
heavily to political and financial factors. We believe
though, that the simple fact that an ISP is proficient in providing
intra-domain tunnels and handling the traffic matrix for his own
network~\cite{HotInterconnects}, combined with the smart routing that the
CXP can perform end-to-end, is a strong basis for the deployment of such models.

\textbf{Preliminary Feasibility Analysis.}
%
We claim that an architecture based on ISP-IXP-CXP collaboration
can work in practice, considering the large number of IP addresses
and the rich path diversity that even a small deployment of CXP anchors can provide.
We illustrate these two properties in Fig.~\ref{fig:IXP_IP_MUF} and
Table~\ref{tab:path-div}, respectively. We use IXP membership data
from~\cite{eu_ix} and build a map of possible pathlets connected to
IXPs. Each pair of IXPs is connected via multiple pathlets traversing the
joint member ASes (a single pathlet per joint AS per IXP pair).
Fig.~\ref{fig:IXP_IP_MUF} depicts the IP address coverage~(\cite{caida}) by IXP members 
(plus their 1-hop customer cone) versus the number of participating IXPs, assuming
an optimal strategy maximizing IP address coverage.
We observe that we can serve over 1 billion IP addresses via a small number
($\sim$5-7) of CXP anchors in well-connected IXPs, if we take into account
only IXP-adjacent prefixes, and over 2 billion IP addresses, if we also consider
the 1-hop customer cone of the IXP members.
This allows an initial deployment of just a few IXPs to serve large parts of the IPv4 address space
and enables incremental adoption. 
Table~\ref{tab:path-div} shows the number of disjoint paths between the
5 largest IXPs, pairwise, based on min-cuts performed on the full extracted pathlet
map derived from~\cite{eu_ix}. 
The very high path diversity results in high path availability,
a competitive marketplace for users, and many choices for selecting
e2e paths tailored to the service requirements (QoS metrics).




\textbf{Acknowledgements.} This work has received funding from the
European Research Council Grant Agreement n. 338402.

\vspace{-15pt}
{\bibliographystyle{acm}
  {\footnotesize
  \setlength{\itemsep}{0pt} 
  \bibliography{refs}
}

\end{document}